\documentclass[a4paper,12pt]{article}
\usepackage{times,cite,amsmath,amsfonts,amsthm,fullpage,graphicx}

\theoremstyle{plain}

\theoremstyle{remark}

\newtheorem{corollary}{Corollary}[section]
\newtheorem{definition}{Definition}[section]

\newtheorem{proposition}{Proposition}[section]

\newtheorem{examps}{Examples}[section]

\newtheorem{lemma}{Lemma}[section]
\newtheorem{remark}{Remark}[section]
\newtheorem{remarks}[remark]{Remarks}

\def\bx{\begin{example}}
\def\ex{\end{example}}
\def\bxs{\begin{examps}. \rm\begin{enumerate}}
\def\exs{\end{enumerate}\end{examps}}
\def\bd{\begin{definition}}
\def\ed{\end{definition}}
\def\bp{\begin{proposition}\rm}
\def\ep{\end{proposition}}
\def\bc{\begin{corollary}}
\def\ec{\end{corollary}}
\def\bl{\begin{lemma}\em}
\def\el{\end{lemma}}
\def\be{\begin{equation}}
\def\ee{\end{equation}}
\def\br{\begin{remark}\rm\small}
\def\er{\end{remark}}
\def\brs{\begin{remarks}.\\ \rm\
\begin{enumerate}}
\def\ers{\end{enumerate}\end{remarks}}
\def\bea{\begin{eqnarray}}
\def\eea{\end{eqnarray}}
\def\bfig{\begin{figure}[!ht]}
\def\efig{\end{figure}}

\def\ln{\mathrm {ln}}

\def\&{&{\hskip -20pt}}

\def\time{{\textsc t}}

\newcount\YDcount\YDcount=0
\def\YDsize{10pt}

\def\YD#1{%
\ifnum#1=0
 \ifnum\YDcount=0 \ifx\varnothing\undefined\emptyset\else\varnothing\fi
 \else\vskip1.4pt\egroup\YDcount=0\fi
\else
 \ifnum\YDcount=0 \YDcount=1\vcenter\bgroup\vskip1pt
 \else\nointerlineskip\fi
 \vbox{\hrule\hbox{\vrule height\YDsize
 \loop\hskip\YDsize\vrule\ifnum\YDcount<#1\advance\YDcount1\repeat}\hrule
 \kern-0.4pt}\expandafter\YD
\fi}

\begin{document}
\baselineskip 16pt
\medskip
\begin{center}
\begin{Large}\fontfamily{cmss}
\fontsize{17pt}{27pt} \selectfont \textbf{Random turn walk on a
half line with creation of particles at the origin}\footnote{This
work has been partially supported by  (1) the European
 Union through the FP6
Marie Curie RTN {\em ENIGMA} (Contract number MRTN-CT-2004-5652)
and the European Science Foundation Program MISGAM
 and by  (2)  the Russian Academy of Science program
``Fundamental Methods in Nonlinear Dynamics",
   RFBR grant No 05-01-00498, and joint RFBR-Consortium E.I.N.S.T.E.IN grant
No 06-01-92054 KE-a}
\end{Large}\\
\bigskip
\begin{large}
 {J.W. van de Leur}$^{\dagger}$\footnote{vdleur@math.uu.nl}
 and
 {A. Yu. Orlov}$^{\star}$\footnote{ orlovs@wave.sio.rssi.ru}
\end{large}
\\
\bigskip
\begin{small}

$^{\dagger}${\em Mathematical Institute,University of Utrecht,\\
P.O. Box 80010, 3508 TA Utrecht,
The Netherlands}\\
\smallskip
$^{\star}$ {\em Nonlinear Wave Processes Laboratory, \\
Oceanology Institute, 36 Nakhimovskii Prospect\\
Moscow 117851, Russia } \\
\end{small}
\end{center}

\begin{center}{\bf Abstract}
\end{center}
\smallskip

\begin{small}
We consider a version of random motion of  hard core particles on
the semi-lattice $ 1,\ 2,\ 3,\ldots$, where in each time instant
one of three possible events occurs, viz., (a) a randomly chosen
particle hops to a free neighboring site, (b) a  particle is
created at the origin (namely, at site $1$) provided that site $1$
is free and (c) a particle is eliminated at the origin (provided
that the site $1$ is occupied). Relations to the BKP equation are
explained. Namely, the tau functions of two different BKP
hierarchies provide generating functions respectively (I) for
transition weights between different particle configurations and
(II) for an important object: a normalization function which plays
the role of the statistical sum for our non-equilibrium system. As
an example we study a model where the hopping rate depends on two
parameters ($r$ and $\beta$). For time $\time\to\infty$ we obtain
the asymptotic configuration of particles obtained from the
initial empty state (the state without particles) and find an
analog of the first order transition at $\beta=1$.
\end{small}
\bigskip

\section{Introduction}

\label{intro} In the famous paper \cite{F} M.Fisher introduced
models of one-dimensional random walk of hard core particles on
the lattice. We shall consider a specific version of the models
that Fisher called random turn walk models. They describe a motion
of particles where at each tick of the clock a randomly chosen
walker takes a random step. In both types of models each site may
be occupied by only one walker at the same time.

In our case we consider a version of this model where particles
move along a semi-line, and where also a particle may be created
at the origin.


{\bf The model}. Consider a set of nodes labelled by positive
integers, where each nearest neighbors are linked by a pair of
opposite arrows. Let us view the nodes $2,\ 3,\ 4,\dots$ as
situated to the right of the origin, the node $1$. An arrow which
starts at a node $i$ and ends at a node $j$ (where $j=i\pm 1$) is
assigned a weight equal to $e^{-{U}_j+{U}_{i}}$. Here
${U}=({U}_{1},{U}_{2},\dots)$ is a set of real numbers.

As an initial state of our dynamical system related to a time
$\time =0$, we place a certain number of hard core particles at
nodes ("hard-core" means that each node may be occupied by at most
one particle), this initial configuration will be denoted by
$\lambda^{(0)}$.

Consider a random motion of the hard-core particles where in each
time instant  one of  three possible events occurs: (a) a randomly
chosen particle hops along any of arrows attached to the
corresponding node (i.e. either to the left or to the right)
 provided that target node is free of particles;
 (b) A  particle is created at the origin (namely, at
site $1$) provided this site is free; (c)  a
particle is eliminated at the origin (provided that site
$1$ is occupied). We shall refer to configurations as {\em
neighboring} ones if they differ by an elementary event from the
list above.

\begin{figure}
\begin{center}
\includegraphics[scale=0.4]{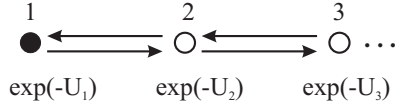}
\caption{Configuration for $\time =1$ in case
$\lambda^{(\time=0)}$ is the empty configuration. Black ball
serves for a particle, white balls for free sites} \label{time=1}
\end{center}
\end{figure}

For instance, the empty configuration is an
neighboring one to the configuration shown on figure
\ref{time=1}, because the last is obtained by the event (b) from
the configuration without particles.

The probability of each elementary (i.e. which occurs in one time
instant) event is proportional to a {\em weight} (or, a {\em
rate}) of the
event which we assign as follows: \\
(a)  For the hop  of a particle along an arrow $i\to j$ (where
$j=i\pm
1$)  the weight  is given by $e^{-{U}_j+{U}_{i}}$.
The weight $r(j):=e^{-{U}_j+{U}_{i}},\;j=2,3,\dots$ will be called (right) hopping rate;\\
(b) For the birth process  the weight is
$\frac{1}{\sqrt 2}e^{-{U}_{1}}$ (this weight will be called birth rate $r(1))$; \\
(c) For the elimination process  the weight is $\frac{1}{\sqrt
2}e^{{U}_{1}} $.

Along the random process in each each time step, say,
$\time\to\time +1$ a configuration $\lambda^{(\time)}$ goes to a
neighboring configuration $\lambda^{(\time +1)}$ with the
following probability:
 \be\label{step-prob}
p_{\lambda^{(\time)}\to\lambda^{(\time
+1)}}=\frac{W_{\lambda^{(\time)}\to\lambda^{(\time +1)}}}{\sum_\mu
W_{\lambda^{(\time)}\to\mu} }\, ,
 \ee
where $W_{\lambda^{(\time)}\to\mu}$ is the weight of the
elementary event which creates (in one time step) a configuration $\mu$ from the
configuration $\lambda^{(\time)} $, the sum in the denominator
ranges over all neighboring configurations. For instance, in case
the initial configuration is the empty configuration, then by
rules (a)-(c) with unit probability we obtain the configuration
depicted in figure \ref{time=1}.

Each set of configurations
$\lambda^{(0)},\lambda^{(1)},\dots,\lambda^{(\time)}$ where each
pair $\lambda^{(i)},\lambda^{(i+1)}$ is a pair of neighboring
configurations will be referred to as a {\em path} of duration
$\time$ which starts from the configuration $\lambda^{(0)}$ and
ends at the configuration $\lambda^{(\time)}$. The {\em weight of
the path}  is defined as the product of the weights of all
elementary events along the path,
$W_{\lambda^{(0)}\to\lambda^{(1)}}W_{\lambda^{(1)}\to\lambda^{(2)}}
\cdots
W_{\lambda^{(\time-1)}\to\lambda^{(\time)}}$. The {\em transition
weight between configurations} $\lambda^{(0)}$ and
$\lambda^{(\time)}$ is defined as the sum of weights of all paths
of  duration $\time$ starting from $\lambda^{(0)}$ ending at
$\lambda^{(\time)}$ and will be denoted by
$W_{\lambda^{(0)}\to\lambda^{(\time)}}(\time)$.

Then, it is easy to see that the probability to come from a
configuration $\lambda^{(0)}$ to a given configuration
$\lambda^{(\time)}$ in $\time$ steps is the ratio
 \be\label{prob} p_{\lambda^{(0)}\to\lambda^{(\time)}}=
\frac{W_{\lambda^{(0)}\to\lambda^{(\time)}}(\time)}{Z_{\lambda^{(0)}}(\time)}
 \ee
where the denominator is  a normalization function
 \be\label{norm-funct}
Z_{\lambda^{(0)}}(\time)=\sum_\mu W_{\lambda^{(0)}\to\mu}(\time)
 \ee

In what follows we shall omit superscripts for configurations. We
shall not need intermediate configurations which were introduced
for better explanation of our model.

Our letter is arranged as follows. In an introductory part to
section \ref{Fock} we introduce our tools: neutral fermions and
the related Fock space, and quadratic operators
(\ref{B1})-(\ref{B-1}) which depends on a given set
$U=(U_1,U_2,\dots)$. In subsection \ref{Stochastic} for arbitrary
set $U$ we shall obtain an explicit expression for the probability
to achieve a given configuration in $\time$ steps in case the
initial configuration is the empty state (the state without
particles), see formulae (\ref{transweight}),
(\ref{discrete-integral}) and (\ref{probability-vac-lambda}). For
an arbitrary chosen external potential $U$ it is impossible to
obtain an asymptotic limit in formulae (\ref{discrete-integral})
and (\ref{probability-vac-lambda}) in the large time limit. In
subsection (\ref{sub2.2}) we specify rate $r(n)\; n=1,2,\dots$ by
formula (\ref{r-n-beta}), now the rate depends on two parameters:
a constant denoted by $r$ and an exponent $\beta$. In this case we
present the asymptotic formulae for the density of particles,
$\sigma(n)$, see (\ref{arcsinus-density-uBB}) and (\ref{T-R}). We
shall show that for a given and large enough $\time$ all
characteristics of the asymptotic configuration of particles:  its
size, the number of involved particles, the center mass e t.c.
undergo a jump at $\beta=1$ (when $\time\to\infty$). We shall show
that the normalization function
$Z_{\lambda^{(0)}=0}(\time)=Z(r,\beta,\time)$ also has a jump. In
our problem this function plays a role quite similar to the role
of the partition function in statistical physics; in this sense in
our model we may treat this jump  at $\beta=1$ as a first order
phase transition. The expression for probability in different
regions of the parameter $\beta$ is given by
(\ref{asymptotic-prob}) and (\ref{f}). Let us note that if we fix
$\beta=1$, then, our model turns out to be a discrete time version
of the model called asymmetric simple exclusion process (ASEP),
now the constant $r$ plays a role of asymmetry parameter. In  two
parts of subsection \ref{Relation to the BKP} we shall link
respectively transition weights (in item I) and the normalization
function (see
 item II) with tau functions of two different BKP, in the item II
the key role  is the relation  (\ref{U-t}) which relates rates
$r(n)$ to the BKP higher times.

Let us notice that a wide usage of the free fermion approach to
random partitions and certain random processes was presented by
Andrei Okounkov in a series of papers, in particular  see
\cite{Okounkov}. Our approach is different and based on
(\ref{evolution-of-Fock-v}). It is also different from the
approach invented by H.Spohn and K-H.Gwa in \cite{GS2} for the
study of (the continues time) ASEP where spin system was used for
the coding of particle configurations. In this approach the state
spin up codes the filled stated, spin down the empty state. Via
Jordan-Wigner tranform this spin system may be related to
fermionic one. However in that case a quadric fermionic
Hamiltoinian is used to describe the stochastic dynamics of
particles which was identified with Hamiltonian dynamics of
quantum spin (or, of nonlinear fermionic) system where the (real)
wave function yields the probability distribution for the
stochastic process. Following \cite{rand} we use free fermions and
the answer for the probability is given by the ratio of two
factors (\ref{probability-vac-lambda}) quite similar to what we
have in thermodynamics where the probability of a state is the
ratio of the weight of the state and a normalization function (the
partition function).

The main part of this work was reported by one of the authors
(J.W.L) on the workshop "Random and integrable models in
mathematics and physics" in Brussel,  September 11-15, 2007.

\section{Fock vectors and configurations of hard core
particles}\label{Fock}

{\bf Neutral fermions}. In what follows we shall need neutral
fermions, $\{ \phi_n\}_{n\in {\bf Z} }$, as introduced by  E. Date, M. Jimbo, M. Kashiwara
and T. Miwa in \cite{DJKM'}, defined by the following property:
\begin{equation}\label{canonical}
    [\phi_n,\phi_m]_+=(-1)^n\delta_{n,-m},
\end{equation}
where $[\ ,\ ]_+$ denotes the anticommutator. In particular,
$\left(\phi_0\right)^2=\frac12$. Next we define an action on vacuum states
 by
 \be\label{phi-on-vacuum}
\phi_n|0\rangle =0, \qquad \langle 0|\phi_{-n}=0,\qquad  n<0,
\qquad \phi_0|0\rangle=\frac{1}{\sqrt 2}|0\rangle, \qquad \langle
0|\phi_0=\frac{1}{\sqrt 2}\langle 0|,
 \ee
 Note that the definition of these "Fock spaces" is different from the usual one which was
introduced in \cite{DJKM'}.
 We follow \cite{KvdLbispec}, where the action of $\phi_0$ (which is not a number) is
 different.
See Appendix for some more details.


The basis of the corresponding right and left Fock spaces are
formed by vectors
 \be\label{Fock-basis-1}
|\lambda\rangle:=\phi_{\lambda_1}\cdots
\phi_{\lambda_N}|0\rangle
 \ee
and by
 \be\label{Fock-basis-3}
\langle \lambda|:=(-1)^{|\lambda|}\langle
0|\phi_{-\lambda_N}\cdots \phi_{-\lambda_1} \, ,
 \ee
where
 \be
\lambda_1>\cdots>\lambda_N>0
 \ee
 and $|\lambda|=\lambda_1+\cdots +\lambda_N$.

 \;

We have {\em one to one correspondence} between configurations of
hard
 core particles on the lattice $1,\ 2,\ 3,\dots$ and the basis Fock
 vectors.
These configurations are also called Maya diagrams. Namely, the
Maya diagram of the vector $|\lambda\rangle$
 is a set of vertices $1,\ 2,\ 3,\dots$, where each vertex numbered by $\lambda_i$ is
 drawn as the black ball (a hard core particle), all other vertices are white balls.

\;

 As we see, by (\ref{canonical})-(\ref{phi-on-vacuum}) we have
 \be\label{orth}
  \langle \lambda |\mu\rangle
  =\delta_{\lambda,\mu}
 \ee

\;

Consider the following operator
\begin{equation}\label{choice}
B=B_1(U)+ B_{-1}(U)
\end{equation}
where $U=(U_{1},U_{2},\dots )$, is a semi-infinite set of numbers
(which may be also considered as variables), and where
\begin{equation}\label{B1}
B_1(U)= \sum_{i>0}^\infty (-1)^{i+1}\phi_i \phi_{1-i}
e^{-U_i+U_{i-1}}=\phi_1 \phi_{0} e^{-U_{1}}-\phi_{2} \phi_{-1}
e^{-U_{2}+U_{1}}+\phi_{3} \phi_{-2} e^{-U_{3}+U_{2}}-\cdots
\end{equation}
and
\begin{equation}\label{B-1}
B_{-1}(U)= \sum_{i\ge 0}^\infty (-1)^{i+1}\phi_i \phi_{-1-i}
e^{-U_i+U_{i+1}}=-\phi_0 \phi_{-1} e^{U_{1}}+\phi_{1} \phi_{-2}
e^{-U_{1}+U_{2}}-\phi_{2} \phi_{-3} e^{-U_{2}+U_{3}}-\cdots
\end{equation}

It is straightforward to check the  relations
 \be\label{B-vac}
B_{-1}B_1-B_1B_{-1}=\frac 12 \ ,\qquad B_{-1}|0\rangle=0=\langle
0|B_1 \,  ,
 \ee
which we shall need soon for the calculation of (\ref{decB}). At
last let us note that  for different purpose the operators $B_1$
and $B_{-1}$ were used in \cite{Q} (namely, to construct  examples
of multivariable hypergeometric functions which are also
multisoliton BKP tau functions \cite{nff}).

\subsection{Stochastic system: description via fermions}\label{Stochastic}

{The birth-death of particles at the origin and their diffusive
motion may be described as follows.}

A sequence of Fock vectors
\begin{equation}\label{evolution-of-Fock-v}
|\lambda'\rangle \rightarrow \left(B_1(U)+
B_{-1}(U)\right)|\lambda'\rangle \rightarrow \cdots \rightarrow
\left(B_1(U)+ B_{-1}(U)\right)^{\time}|\lambda'\rangle
\rightarrow \cdots
\end{equation}
describes an evolution of the initial (basis) Fock vector
$|\lambda' \rangle $ - where the variable $\time =0,1,2,\dots$
plays a role of discrete time - to linear combinations of
different basis Fock vectors. Due to the correspondence between
basis Fock vectors and configurations of hard core particles, this
evolution may be interpreted as the random process described in
section \ref{intro}, where each time step is numbered by $\time$.
The details will be presented below.

Let us notice that each basis Fock vector
\begin{equation}\label{}
|\lambda\rangle:=\phi_{\lambda_1} \cdots\phi_{\lambda_N} |0\rangle
\end{equation}
where $\lambda$ is a strict partition which  is
$\lambda=(\lambda_1,\lambda_2,\dots,\lambda_{N})$, that is in one-to-one correspondence
with a configuration of particles, located in the nodes with numbers
$\lambda_1>\dots >\lambda_N> 0$.
This is the reason, why we have chosen (\ref{phi-on-vacuum}) as definition of the Fock space
and not the one of \cite{DJKM'}.
For the Fock space introduced in that article this is not the case.

We are interested in the discrete-time version of this random
process which is given by (\ref{evolution-of-Fock-v}), where each
time step is numbered by $\time$.

As we see via (\ref{canonical}) and (\ref{phi-on-vacuum}), the
first term in the right hand side of (\ref{B1}) describes the
creation of  a hard-core particle located at the node
number $1$ (the origin)  (provided that this
node is free). Since $\phi_0|0\rangle=\frac{1}{\sqrt 2}|0\rangle$ we assign the weight
$\frac{1}{\sqrt 2}e^{-U_1}$ to this creation process. Each other term of $B_1$ describes a
hop to the right. Similarly, the first term of $B_{-1}$ in the right hand
side of (\ref{B-1}) describes the elimination of a
particle located at the node $1$ (provided that this node is
occupied). Since again $\phi_0|0\rangle=\frac{1}{\sqrt 2}|0\rangle$ we assign the weight
$\frac{1}{\sqrt 2}e^{U_1}$ to each elimination process. Other terms of $B_{-1}$
describes the hops to the left (backward motion in the direction
of the origin).

It is not difficult to see that
\begin{equation}\label{W}
W_{\lambda'\to\lambda}(U;\time):=\langle \lambda| \left(B_1(U)+
B_{-1}(U)\right)^{\time}|\lambda'\rangle
\end{equation}
is a sum of weights of paths over all paths of duration $\time$
starting at the configuration $\lambda'$ and ending at the
configuration $\lambda$, and, therefore, yields the transition
weight of the $\time$-step random process which describes a
transition from an initial configuration of the hard-core
particles described by coordinates $\lambda_1',\dots
,\lambda_{N'}'$ to a target configuration $\lambda_1,\dots
,\lambda_N$ defined in the Introduction. As the number of
particles does not have to be conserved along the process, $N$ is
not necessarily equal to $N'$. Let us mark that for each path from
a configuration $\lambda'$ to a configuration $\lambda$ of a
duration $\time$ we have
 \be\label{particle-number}
N-N'= n_+ -n_-
 \ee
 \be\label{time-hops-cr-el}
\time= j_+ +j_- + n_+ + n_-
 \ee
 \be\label{weight-hops-cr-el}
|\lambda|-|\lambda'|=j_+ -j_- + n_+ - n_-
 \ee
 where $j_+$ is the number of hops to the left during the time interval $\time$, $j_-$ is the
 number of hops to the right,  $n_+$ is the number  of creations of
 a particle and $n_-$ is the number  of eliminations of
  particles at the node $1$.
 Let us notice that from (\ref{time-hops-cr-el}) and
 (\ref{weight-hops-cr-el}) it follows that $\time$ and
 $|\lambda|-|\lambda'|$ have the same parity.

\;

Consider the case $\lambda'=0$ and look for the weight of the
process which transports the initial empty state (there are no
particles at all) to a given configuration $\lambda$ in
$\textsc{t}$ steps
\begin{equation}\label{fermtauB0}
W_{0\to\lambda}(U;\time):=\langle \lambda| \left(B_1(U)+
B_{-1}(U)\right)^{\time}|0\rangle
\end{equation}
In order to evaluate the right hand side we need the following
formulae
\begin{equation}\label{decB}
e^{z( B_{-1}+ B_{1})}=e^{\frac{ z^2}{4}}e^{z B_{1}}e^{z
B_{-1}},\quad e^{z B_{-1}}|0\rangle =|0\rangle \ ,
\end{equation}
(the first equation follows from the Baker-Campbell-Hausdorff
formula and from (\ref{B-vac})), and also the formula
\begin{equation}\label{lemmaQ}
\langle \lambda|e^{zB_{1}(U)}|0\rangle
  = 2^{-\frac N2}z^{|\lambda|}
 e^{-\sum_{i=1}^N U_{\lambda_i}}\prod_{i=1}^N \frac
{1}{\lambda_i!}\prod^N_{i<j}\frac{\lambda_i-\lambda_j}
{\lambda_i+\lambda_j} \ ,
\end{equation}
 see \cite{Q},\cite{nff}. Here $|\lambda|:=\lambda_1+\cdots
 +\lambda_N$ is the weight of $\lambda$. In our case $|\lambda|/N$
 is interpreted as the location of the center of mass of the hard core
 particles.

\;

Now using
(\ref{decB}), we obtain
 \be
\label{23}
\langle \lambda| e^{zB_1(U)+ zB_{-1}(U)}|0\rangle=e^{\frac{
z^2}{4}}\langle \lambda| e^{zB_1(U)}e^{ zB_{-1}(U)}|0\rangle
=e^{\frac{ z^2}{4}}\langle \lambda| e^{zB_1(U)}|0\rangle
 \ee
If we develop the left-hand side and the right-hand side of (\ref{23})
in powers of the variable $z$, we obtain the following formula for the
transition weight from the vacuum state to a state $\lambda$ in a
time duration $\time $:
 \be
\langle \lambda| (B_1(U)+ B_{-1}(U))^\time|0\rangle=W_{0\to
\lambda}(\time)=\frac{\time!}{m!}\left(\frac{1 }{4}\right)^m
 2^{-\frac N2}
 e^{-\sum_{i=1}^NU_{\lambda_i}}\prod_{i=1}^N \frac
{1}{\lambda_i!}\prod^N_{i<j}\frac{\lambda_i-\lambda_j}
{\lambda_i+\lambda_j},
 \ee
where the right hand side is non-vanishing only if the relation
 \be
\time=|\lambda|+2m,\quad m=0,1,2,...
 \ee
 is valid. By (\ref{time-hops-cr-el}) and
 (\ref{weight-hops-cr-el}) we obtain the meaning of $m$, viz.:
  \be
m=j_- +n_-
  \ee

Thus
\be
\label{transweight}
W_{0\to\lambda}(U;\time)=
\begin{cases}0&\mbox{ iff }\ \time-|\lambda|\quad \mbox{odd,}\\
\frac{\time !}{\left(\frac{\time-|\lambda|}2 \right)!}
 2^{|\lambda|-\time -\frac N2}
 e^{-\sum_{i=1}^NU_{\lambda_i}}\prod_{i=1}^N \frac
{1}{\lambda_i!}\prod^N_{i<j}\frac{\lambda_i-\lambda_j}
{\lambda_i+\lambda_j}&\mbox{ iff }\ \time-|\lambda|\quad
\mbox{even}
\end{cases}
\ee

The normalization function, counting weights for all possible
target configurations, which may be achieved in the time duration
$\time$,  is
\begin{equation}\label{normalization}
Z(U;\time):=\sum_{N=0}^\infty \sum_{\lambda_1>\cdots
>\lambda_{N}> 0}\langle \lambda| ( B_{-1}+
B_{1})^{\time}|0\rangle
\end{equation}
\begin{equation}\label{normalization-2}
 =\time!\sum_{N=0}^\infty 2^{-\frac N2}
\sum_{\lambda_1>\cdots >\lambda_{N}> 0 \atop \time-|\lambda| \;
even} \frac{2^{|\lambda|-\time}}{\Gamma(\frac
{\time-|\lambda|}{2}+1)}
 \prod_{i=1}^N \frac
{e^{-U_{\lambda_i}}}{\lambda_i!}\prod^N_{i<j}\frac{\lambda_i-\lambda_j}
{\lambda_i+\lambda_j}
\end{equation}
Note that due to the Gamma function this sum is finite.

The term
corresponding to $N=0$ gives in the summation the term $\langle 0|
( B_{-1}+ B_{1})^{\time}|0\rangle$.

\;

Let us note that

(1) $|\lambda|=0$  corresponds to the returning to the initial
position. This only occurs when the number of time instants is
even, $\time =2m$,
 \be\label{W-0-0}
W_{0\to 0}(\time)=\frac{\time !}{\left(
\frac{\time}{2}\right)!}2^{-\time} = 2^{-{\frac \time 2}}(\time
-1)!!
 \ee

(2) Now consider the next case where the final configuration is
the one-particle one. Now $\lambda$ is a number which denote the
coordinate of this particle. For simplicity we consider
$r(n)=r=const$. Then, given large enough $\time$, by Stirling's
formula we can present the transition weight as follows
\[
W_{0\to\lambda} =\frac{2^{\lambda -\time
 -\frac 12}}
 {\lambda !}
 \frac{\time !}{\left(\frac{\time -
\lambda}{2} \right)!}
 e^{-U_{\lambda}} =W_{0\to{\tilde\lambda}}\exp \left(
-\frac{(\lambda-{\tilde\lambda})^2}{2\sqrt{\time}}\right),
\]
where
\[
\quad W_{0\to{\tilde\lambda}}=\frac{\time !2^{{\tilde\lambda}
-\time
 -\frac 12}}
 {\Gamma({\tilde\lambda} +1)\Gamma\left(\frac{\time -  {\tilde\lambda} }{2}+1 \right)}
  r^{{\tilde\lambda} -1}e^{-U_1},\quad {\tilde\lambda}_1=r\sqrt{2\time}-r^2-\frac 12+ O(\time^{-\frac
12})
\]
The formula resembles formula for the Brownian motion (here the
variance is given by $\time^{\frac14}$).
 At last we note, that in
case the rate $r$ depends on site, then, in a wide class of rates
in large $\time $ limit ${\tilde\lambda}_1$ may be evaluated as
the solution of
${\tilde\lambda}_1=r({\tilde\lambda}_1)\sqrt{2\time}$. For
instance, for Gauss potential, $U_n=\frac12 n^2$, one obtains
${\tilde\lambda}_1\sim \log(2\time)$.

(3) For given $\lambda$ in large $\time$ limit (this means that
$\time \gg |\lambda|$) by the Stirling's approximation we have
 \be\label{W-stirling}
W_{0\to \lambda}(\time) = W_{0\to
0}(\time)\time^{\frac{|\lambda|}{2}} 2^{-\frac N2}e^{-\frac
{|\lambda|}{2}}2^{|\lambda|}e^{o(\time^0)}e^{-E_\lambda},
 \ee
where
 \be
\label{electrostatic}
 E_\lambda=-\log \; \prod_{i=1}^N \frac
{e^{-U_{\lambda_i}}}{\lambda_i!}\prod^N_{i<j}\frac{\lambda_i-\lambda_j}
{\lambda_i+\lambda_j}
 \ee
means an electrostatic energy of Coulomb particles (placed in an
external field) which are attracted by their image. We see that in
the large time limit the weight of a configuration increases with
$|\lambda|$, and for given $\time$ and $|\lambda|$ depends only on
the $E_\lambda$.

(4)  The case $\time=|\lambda |=n_+ +j_+$  corresponds to the
non-stop creation + forward motion processes. Then
 \be\label{TASEP}
W_{0\to \lambda}(\time)={\time !}
 2^{-\frac N2}
 e^{-\sum_{i=1}^NU_{\lambda_i}}\prod_{i=1}^N \frac
{1}{\lambda_i!}\prod^N_{i<j}\frac{\lambda_i-\lambda_j}
{\lambda_i+\lambda_j}
 \ee
In case the potential is a rapidly decreasing functions
$U_{n-1}>>U_{n}$ (and, therefore, left hopping rates are much
larger than right hopping rates), then the configurations where
$\time=|\lambda |$ are dominant in the sum for normalization
function.

Let us note that up without the factor $2^{-\frac
N2}e^{-\sum_{i=1}^NU_{\lambda_i}}$ the number $W_{0\to
\lambda}(\time)$ is equal to the number of shifted standard
tableau of shape $\lambda$, see \cite{Mac}, that is the number of
ways the Young diagram of the strict partition $\lambda$ may be
created by adding box by box to the empty partition in a way that
on each step we have the diagram of a strict partition.

Finally, one can get rid of the restriction $\lambda_1>\cdots >\lambda_N$
in the summation (\ref{normalization-2}) rewriting it as a sum
over all non-negative integers $\lambda_1,\cdots ,\lambda_N$:
 \be\label{discrete-integral}
 Z(U;\time)={\time}!\sum_{N=0}^\infty \frac{2^{-\frac N2}}{N!}
\sum_{\lambda_1,\cdots ,\lambda_{N}> 0 \atop \time-|\lambda| \;
even} \frac{2^{|\lambda|-\time}}{\Gamma(\frac
{\textsc{t}-|\lambda|}{2}+1)}
 \prod_{i=1}^N \frac
{e^{-U_{\lambda_i}}}{\lambda_i!}\prod^N_{i<j}\left|\frac{\lambda_i-\lambda_j}
{\lambda_i+\lambda_j}\right|
 \ee

 The probability to come to a configuration $\lambda$ in $\time$ steps starting
 from the vacuum one is given by
 \be\label{probability-vac-lambda}
p_{0\to\lambda}(U;\time)=\frac{W_{0\to\lambda}(U;\time)}{Z(U;\time)}
 \ee

 In the present paper we do not write down the expression for the
 probability $p_{\nu\to\lambda}(U;\time)$ because it is rather
 cumbersome and contains the so-called skew projective Schur
 functions.

\subsection{Asymptotic configuration of the particles in $\time\to\infty$ limit}
\label{sub2.2}

One may ask what configuration is obtained from the empty
configuration (vacuum) configuration due to the
creation/annihlation processes at the edge of the lattice and to
the hops of the particles in the large $\time$ limit. (Such
configuration will be called asymptotic one and denote as
$\lambda(\time)$). To find it we should find the largest term in
(\ref{discrete-integral}). Let us do it via saddle point method.
First, in the large $\time$ limit it is reasonable to introduce
the density of particles $\sigma (u)$, where $u=\frac \lambda R $
where $R$ is the size of configuration. The number of particles is
 \be
\label{sigma-N}
 N=R\int_0^1 \sigma(u)du
 \ee

Let us note that the density interpolates between full package state ($\sigma(u) = 1$),
and empty state ($\sigma(u) = 0$):
 \be
\label{+2} 0\le\sigma(u)\le 1
 \ee

Let us find the asymptotic configuration in the case when the
creation rate is $r(1):=\frac{e^{-U_1}}{\sqrt{2}}=r$ and the
external potential is
 \[
U_n=-n\log r +(\beta -1) \log n!\; ,\quad n=2,3,\dots
 \]
which means that creation and (the right) hopping rates are as
follows
 \be\label{r-n-beta}
r(n)=rn^{1-\beta}\;,\quad n=1,2,3,\dots
 \ee
 $\beta>1$ describes a locking potential while in $\beta<1$ case
the potential try to drive particles to the right from the origin.
The case $\beta=1$ may be considered as a discrete time version of
the so-called asymmetric simple exclusion process (ASEP) on the
half-line, now, the parameter $r$ being an asymmetry parameter.

Let notice that formally the point $n=n_*=r^{\frac{1}{1-\beta}}$
is a point of an extremum of the potential $U_n$ where the left
and the right hopping rates are equal.

It means we want to find the configuration
$\lambda=\lambda(\time)$ where for given $\time$ the weight is as follows
  \be
W_{0\to\lambda}(r,\beta;\time)=
\time!\frac{2^{|\lambda|-\time}}{\left(\frac{\time-|\lambda|}2
\right)!}
 2^{-\frac N2}e^{ - E_\lambda(r,\beta)}
 \ee
here $\time-|\lambda|$ is even, and the''electrostatic energy`` of the configuration is
 \be\label{E-r} E_\lambda(r,\beta)= \; - \; \log \; \left( r^{|\lambda|}
 \prod_{i=1}^N \frac
{1}{(\lambda_i!)^\beta}\prod^N_{i<j}\frac{\lambda_i-\lambda_j}
{\lambda_i+\lambda_j}\right)
 \ee

In the continues limit in a standard way one can write the saddle
point equation  for sum $Z(\time)$ which will define the density
function $\sigma$ in the large time limit, $\time\to\infty$. For
$\lambda\in (0,R)$ we get \be\label{int-sing-eq} \log
\frac{r}{\lambda^{\beta}} +P \int_0^R
\frac{\sigma(xR^{-1})dx}{\lambda-x}- P \int_0^R
\frac{\sigma(xR^{-1})dx}{\lambda+x}+\frac12 \log 2\left(
\time-\int_{0}^{R}x\sigma(x R^{-1})dx \right)=0
 \ee
where $P\int$ stands for the principal value. This equation is to
be solved by a standard method \cite{Gakhov}, see Appendix.
 For $0<\beta<2$ we find
 \be\label{arcsinus-density-uBB}
\sigma(\lambda R^{-1})=\frac{\beta }{\pi }\arccos \frac
{\lambda}{R},\quad \lambda\in [0,R]
 \ee
(The validity of (\ref{arcsinus-density-uBB}) may
 be verified by substitution in (\ref{int-sing-eq})).
The constraint (\ref{+2}) causes the restriction
  $0<\beta < 2$ for our solution in formula (\ref{arcsinus-density-uBB}).

 The weight of such configuration is
 \be\label{area-1}
|\lambda(\time)|=R^2\int_0^1 u\sigma(u)du=\frac{\beta R^2}{8}
 \ee
and after substitution into the logarithmic term in
(\ref{int-sing-eq}) we obtain a relation of $R$ to $\time$ as
follows
 \be\label{T-R}
\time = \frac{\beta}{8}R^2 + \frac {2^{2-2\beta}}{8r^2}R^{2\beta}
 \ee
  Let us make a remark. In the approximation
we consider one can replace sums by integrals only up to terms of
the order $O(R^0)$. It results to a fact that $r$ in the relation
(\ref{T-R}) should be replaced by an effective (yet undefined)
hopping rate $r_{\mbox{eff}}=r/r_0$, where $r_0$ is of order
$e^{O(R^0)}$. Below by $r$ we will imply this effective hopping
rate. Arguments that $r_{\mbox{eff}}=r$ will be considered
separately.

Equation (\ref{T-R}) shows
 that in the large $\time $ limit the dependence of $R$ on $\time$ is different in
 regions $0<\beta < 1$, $\beta =1 $ and $1<\beta <2$. Let us introduce
\[
 R_*=2(\beta r^2)^{\frac{1}{2(\beta-1)}}, \qquad
 \time_*=\beta(\beta r^2)^{\frac{1}{\beta-1}}
\]
As we see in the region when $R=R_*$ the terms in the r.h.s. of
(\ref{T-R}) are equal. Below we imply that $\time >> \time_* $.
 In the large $\time$ limit as we see from (\ref{T-R})
 \be \label{R-asympt}
R=R(\beta,\time)=
\begin{cases}
\sqrt\frac{8\time}{\beta} &\mbox{ if }\; 0<\beta < 1  \\
\sqrt{\frac{8\time}{1+r^{-2}}}  &\mbox{ if }\; \beta =1 \\
 2\left( 2 r^2 \time \right)^{\frac{1}{2\beta}} &\mbox{ if}\;  1<\beta
 <2
\end{cases}
 \ee
Thus, when $0\le\beta <1 $ the size of the asymptotic
configuration is proportional to $\sqrt{\time}$, while in the
vicinity of $\beta=2$ we have the forth root behavior. As we see
the discontinuity appears at $\beta=1$ in the large $\time$ limit.
The same behavior has the number of particles in the asymptotic
configuration which is proportional to the size of the
configuration:
 \be\label{asympt-N}
N(r,\beta,\time) = R\int_0^1 \left(\frac {\beta}{\pi} \arccos u
\right)du =\frac {\beta R(\beta,\time)}{\pi }
 \ee
The weight of the asymptotic configuration (\ref{area-1}) for
large enough $\time$   is
 \be\label{area}
|\lambda(\time)|=\frac{\beta R^2}{8} \approx
\begin{cases}
\time +O\left( \time^{\beta}\right)&\mbox{ if }\ 0<\beta < 1  \\
\frac{\time}{1+r^{-2}}  &\mbox{ if }\ \beta =1 \\
\frac{\beta}{2}
 \left( 2 r^2 \time \right)^{\frac{1}{\beta }} &\mbox{ if}\
 1<\beta
 <2
\end{cases}
 \ee

\begin{figure}
\begin{center}
\includegraphics[scale=0.4]{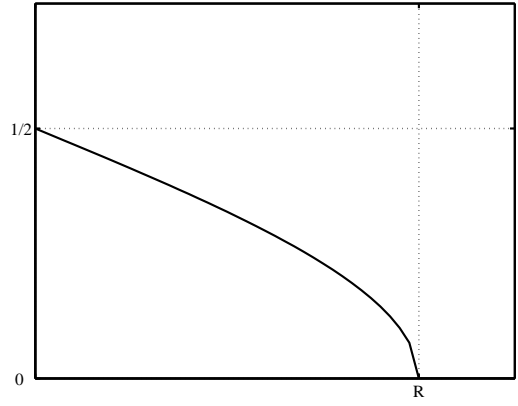}
\caption{ The evolution of the empty configuration in $\time \to
\infty$ limit for $\beta=1$. Asymptotic density of particles.}
\label{asymptotic}
\end{center}
\end{figure}

Notice that the center mass of the configuration also has a jump
at $\beta=1$:
 \be\label{c.m.-asympt}
 \frac{|\lambda(\time)|}{N(r,\beta,\time)}\approx
\begin{cases}
\pi\sqrt{\frac{\time}{8\beta}} &\mbox{ if }\ 0<\beta < 1  \\
\pi\sqrt{\frac{\time}{8(1+r^{-2})} } &\mbox{ if }\ \beta =1 \\
 \frac\pi 4
 \left( 2 r^2 \time \right)^{\frac{1}{2\beta }} &\mbox{ if}\ 1<\beta < 2
\end{cases}
 \ee

For large enough $\time$, the number
$m(\time)=j_-(\time)+n_-(\time)$ depends on region as follows
 \be\label{m-asympt}
 m(\time)=\frac 12 (\time-|\lambda|) =\frac 12 \time-\frac{\beta R^2}{16}=
\begin{cases}
O\left( \time^{\beta}\right) &\mbox{ if }\ 0<\beta < 1  \\
\frac{\time}{2(1+r^{2})}  &\mbox{ if }\ \beta =1 \\
 \frac\time 2 - \frac{\beta}{4}
 \left( 2 r^2 \time \right)^{\frac{1}{\beta }} &\mbox{ if}\ 1<\beta < 2
\end{cases}
 \ee
In (\ref{area}) and (\ref{m-asympt}) we keep terms which we shall
use in evaluations below.

 For large enough $\time$, the electrostatic energy (\ref{E-r}) of $\lambda(\time)$ is
 \[
E_{\lambda(\time)}(r,\beta)=-|\lambda(\time)|\;\ln\; r\; +
 \]
 \be
\left(\beta R^2\int_0^1{\sigma}(u)\left(u\ln(uR)-u \right)du
\right) -\frac 12  R^2\int_0^1\int_0^1
{\sigma}(u){\sigma}(u')\ln\frac{u-u'}{u+u'}dudu'
 \ee
 \[
 = |\lambda(\time)|\left(\beta\;\ln\; R -\ln\; r \right)-R^2\beta^2 A
 \]
where
 \be
\beta^2 A=-\beta\int_0^1{\sigma}(u)\left(u\ln u-u \right)du+
  \int_0^1 du'\int_0^u
{\sigma}(u){\sigma}(u')\ln\frac{u-u'}{u+u'}du=\beta^2\left(\frac{1}{16}+\frac{\ln\;
2}{8} \right)
 \ee
 see Appendix \ref{Two useful integrals} which yields
 $A = \frac{1}{16}+\frac{\ln\; 2}{8} $.

By (\ref{R-asympt}) and (\ref{area}) we obtain

 \be
E_{\lambda(\time)}(r,\beta)  \approx
\begin{cases}
\frac{\beta}{2} \time \ln \time  -
\time\left( \frac{\beta }{2} \ln \frac{\beta }{2} + \frac{\beta }{2} + \ln\; r  \right) &\mbox{ if }\ 0<\beta < 1  \\
\frac 12 \frac{\time\ln \time}{1+r^{-2}}+
\frac{\time}{1+r^{-2}}\left(\frac 12\ln \frac{8}{1+r^{-2}} -8A - \ln\; r \right) &\mbox{ if }\ \beta =1 \\
\frac{\beta}{4}
 \left(2 r^2 \time \right)^{\frac{1}{\beta}}
\left(\ln \time + (1+2\beta )\ln 2 -{16}\beta A  \right) &\mbox{
if}\; 1 < \beta < 2
\end{cases}
 \ee

Using Stirling's approximation to evaluate $m(\time)!$, for
$\time$ large, we obtain the weight of the process as follows
 \be
W_{0\to\lambda(\time)}(r,\beta;\time)=\time!
\frac{2^{-2m(\time)}}{m(\time)!}
 2^{-\frac {N(\time)}2}e^{-E_{\lambda(\time)}(r,\beta)}
 \ee
 \be
 =
\begin{cases}
\time!e^{-\frac{\beta }{2}\time\ln\time +
\time\left(\frac{\beta}{2}\ln \frac{\beta}{2} +\frac{\beta}{2} +
\ln r \right) + O(\sqrt{\time})}
 &\mbox{ if }\ 0<\beta < 1  \\
\time!e^{ -\frac{\time}{2}\ln\;{\time}+\time \{ \frac{\ln\;
2(1+r^2)}{2(1+r^2)}+\frac{\ln\; 2(1+r^{-2})}{2(1+r^{-2})}
+\frac{2\ln\; 2 -1 + \ln\; r^2}{2(1+r^{-2})}+\frac{1-\ln\;
4}{2(1+r^2)} \} + O(\sqrt{\time})}
 &\mbox{ if }\ \beta =1 \\
{W_{0\to 0}(\time)}e^{- \frac{\beta}{4}
 \left(2 r^2 \time \right)^{\frac{1}{\beta}}
\left( 1-\beta \right)+O\left(\time^{\frac{1}{2\beta}} \right)}
&\mbox{if}\; 1<\beta <2
\end{cases}
 \ee

Then, as we see the normalization function $Z(\time)$ which
according to the saddle point method has the same leading term in
the large $\time $ limit as $W_{0\to\lambda(\time
)}(r,\beta;\time) $ has a discontuinity at $\beta=1$ which may be
interpreted as a sort of the first kind phase transition in our
non-equilibrium system.

Now we can evaluate the type of asymptotic of the probability to
achieve a given configuration in $\time\to\infty$ steps. We have
$Z(\time)=W_{0\to\lambda(\time)}(\time)e^{O(\ln R)}$, where the
last factor originates from the Gaussian integral around the
saddle point configuration $\lambda(\time)$. Then for $\time \gg
|\lambda|$ we have
 \be\label{asymptotic-prob}
p_{0\to\lambda}(r,\beta,\time) \approx
\frac{W_{0\to\lambda}(r,\beta,\time)}{W_{0\to\lambda(\time)}(r,\beta,\time)}
\approx \time^{\frac{|\lambda|}{2}} e^{-\frac
{|\lambda|}{2}}2^{|\lambda|}e^{-E_\lambda(r,\beta)}e^{\omega(\time,r,\beta)}
 \ee
where $\omega$ does not depend on $\lambda$. (For the enumerator
of (\ref{asymptotic-prob}) we used (\ref{W-stirling}) and
Stirling's approximation for (\ref{W-0-0}).)

The answer depends on the region of $\beta$
  \be\label{f}
e^{\omega(\time,r,\beta)}=
  \begin{cases}
e^{\frac{\beta -1}{2}\time\ln\time - \time\left(\frac{\beta}{2}\ln
\frac{\beta}{2} +\frac{\beta}{2}+\ln
r)\right)+ \frac{\time}{2}\ln(2e)+O(\sqrt{\time})} &\mbox{ if }\ 0<\beta < 1  \\
e^{ -\time \left\{ \frac{\ln\; 2(1+r^2)}{2} +\frac{b}{2\left(1+r^{-2}\right)} \right\}
 + O(\sqrt{\time}) }  &\mbox{ if }\ \beta =1 \\
e^{\frac{\beta}{4}
 \left(2 r^2 \time \right)^{\frac{1}{\beta}}
\left( 1-\beta \right) +O\left(\time^{\frac{1}{2\beta}} \right)}
 &\mbox{ if}\;
1<\beta <2
\end{cases}
 \ee
where $b= 2\ln\;2-1\approx 0.4$.
 As we see, in each case, in the large $\time$ limit $e^{\omega}$ is vanishing.

 \;

At last let us note the following. As we see in case of a
decreasing potential (or, the same, a increasing rightward hopping
rate), $ 0<\beta < 1 $, the weight of the asymptotic configuration
is equal to $\time$ which means that the asymptotic configuration
is created by only creating events at the origin and rightward
hops, there were no elimination events and backward hops in the
history of this configuration, $j_-=n_-=0$.

 For $\beta < 0$ solution does not exists.  In $\beta\to +0$ limit the number of particles
(\ref{asympt-N}) vanishes, while $|\lambda(\time)|$ is equal to
$\time$. Indeed, the external potential $U_n$ is decreasing so
rapidly that the largest weight has the one particle configuration
 where the particle moves in the ballistic way: it is located at the distance $\time$
 to the origin. When $\beta > 2$ we have a locking potential which forces particles to
  form a sort of a condensate near the origin where all sites are occupied. The size
of the condensed phase is defined by $\beta $. This problem is
 treated by a method similar to the suggested in \cite{Douglas-Kazakov}. Along this way we
can show that the solution is given in terms of elliptic integrals
of the first and third kind. The problem will be considered in
detail in the next version of this paper, where we will show that
the normalization function $Z(\time)$ has singularity as function
of the parameter $\beta$ at $\beta=2$.

\subsection{Relation to the BKP tau function}\label{Relation to the BKP}

The link of the described stochastic system to  integrable
equations is two-fold.

\noindent
(I) {\bf A BKP tau function as generating function for transition
weights $W_{\lambda'\to\lambda}(\time)$}.

Let us consider the following vacuum expectation value
 \be\label{BKP-A}
\tau(s,{\bar s};U,z):=\langle 0| e^{H(s)}e^{zB}
e^{{\bar{H}}({\bar{s}})}|0\rangle
 \ee
 where  $B$ is given by (\ref{choice}) and
 \be\label{Hamiltonians}
H(s)=\frac12 \sum_{n=1,3,5,\dots}\sum_{m\in\mathbb{Z}} s_n
(-1)^{m+1}\phi_m\phi_{-n-m},\quad {\bar H}({\bar
s})=\frac12 \sum_{n=1,3,5,\dots}\sum_{m\in\mathbb{Z}} {\bar s}_n
(-1)^{m+1}\phi_m\phi_{n-m}
  \ee
with $z$, $U=(U_1,U_2,\dots)$  and sets
$s=(s_1,s_3,s_5\dots)$ and ${\bar s}=({\bar s}_1,{\bar s}_3,{\bar
s}_5,\dots)$  parameters. Function $\tau(s,{\bar s};U,z)$
depends on $U=(U_1,U_2,\dots)$ as the operator $B$ depends on
these parameters.

 The hierarchy of of Kadomtsev-Petviashvili equations of
 type B (the BKP hierarchy) was introduced in \cite{DJKM'}.
 As we
 have already mentioned we use its modification suggested in
 \cite{KvdLbispec}.
These is a semi-infinite set of compatible nonlinear differential
equations which may be viewed as a set of  commutative time flows.
It is common to enumerate the BKP equations (flows) by odd
numbers.

Expression (\ref{BKP-A}) where $U$ may be chosen as an arbitrary
set of numbers, provides an example of the BKP tau function
constructed in \cite{KvdLbispec}, where $s=(s_1,s_3,s_5\dots)$ is
the set of the so-called higher times (the times of commutative
flows related to different equations of the BKP hierarchy).
Actually the set
 ${\bar s}=({\bar s}_1,{\bar s}_3,{\bar s}_5,\dots)$ is also related to
 a (second) BKP hierarchy of equations, which is compatible with the
 first one; thus, (\ref{BKP-A}) is an example of the tau function of a
 coupled BKP hierarchy.

Using results of  \cite{You} and doing similar calculations in the
framework of BKP hierarchy constructed in \cite{KvdLbispec} (see
Appendix) one can show that the operator $e^{H(s)}$ applied to the
left vacuum generates all basis left Fock vectors, quite similar
$e^{{\bar H}({\bar s})}$ applied to the right vacuum vector
generates all basis right Fock vectors. Namely, we have the
following left and right coherent states (see also
(\ref{Fock-basis-1}) and (\ref{Fock-basis-3}))
 \be \langle
0|e^{H(s)}=\sum_{N=0}^\infty 2^{-\frac N2} \sum_{\lambda_1>\dots
>\lambda_N > 0}Q_\lambda\left(\frac{s}{2}\right)\langle \lambda|\,
,
 \ee
 \be
 e^{{\bar H}({
s})}|0\rangle =\sum_{N=0}^\infty 2^{-\frac
N2}\sum_{\lambda_1>\dots >\lambda_N
> 0}Q_\lambda\left(\frac{s}{2}\right)|\lambda\rangle
 \ee
where $Q_\lambda\left(\frac{s}{2}\right),\;
\lambda=(\lambda_1,\lambda_2,\dots)$, are known as projective
Schur functions, see \cite{Mac} \footnote{Here we use notations of
\cite{You}. See also \cite{Nimmo} for the relation of the
projective Schur functions to the BKP hierarchy.}, which, in an
appropriate space, form a complete set of weighted polynomials in
the variables $s_1,s_3,\dots$. Using these formulae we obtain that
the BKP tau function (\ref{BKP-A}) is the generating function for
(\ref{W}), namely
  \be\label{doubleQSchur}
\tau(s,{\bar s};U,z)=\sum_{\time=0}^\infty \sum_{N,N'=0}^\infty
2^{-\frac {N}{2}-\frac {N'}{2}} \sum_{{ \lambda_1>\dots >\lambda_N
> 0}\atop{ \lambda_1'>\dots >\lambda_{N'}'> 0} }
\frac{z^{\textsc{t}}}{{\textsc{t}}!}
Q_{\lambda}\left(\frac{s}{2}\right) Q_{\lambda'}\left(\frac{\bar
s}{2}\right) W_{\lambda'\to\lambda}(U;{\textsc{t}})
 \ee

What one obtains in case of a general BKP tau function will be
explained in a more detailed forthcoming paper.

(II) {\bf A tau function of dual BKP as the generating function
for the normalization function $Z(\time)$}

Let us introduce the generating function for the normalization
function (\ref{normalization}) as follows
 \be\label{Zz}
Z(U;z):=\sum_{N=0}^\infty\sum_{\lambda_1>\cdots >\lambda_N>
0}\langle \lambda|e^{zB}|0\rangle=\sum_{\time =0}^\infty
\frac{z^\time}{\time!}Z(U;\time)
 \ee

By (\ref{decB}) and (\ref{lemmaQ}) we have
 \[
Z(U;z)=e^{\frac{ z^2}{4}}\sum_{N=0}^\infty 2^{-\frac
N2}\sum_{\lambda_1>\cdots
>\lambda_N> 0}z^{|\lambda|}
 e^{-\sum_{i=1}^NU_{\lambda_i}}\prod_{i=1}^N \frac
{1}{\lambda_i!}\prod^N_{i<j}\frac{\lambda_i-\lambda_j}
{\lambda_i+\lambda_j}
 \]
 \be\label{Zz-serie}
=e^{\frac{ z^2}{4}}\sum_{N=0}^\infty \frac{2^{-\frac
N2}}{N!}\sum_{\lambda_1,\cdots , \lambda_N> 0} \prod_{i=1}^N \frac
{ e^{\lambda_i\ln
z-U_{\lambda_i}}}{\lambda_i!}\prod^N_{i<j}\left|\frac{\lambda_i-\lambda_j}
{\lambda_i+\lambda_j}\right|
 \ee
Let us mark that the last sum may be compared with the grand
partition function of the log Coulomb gas, compare with
\cite{Forr1}, \cite{LS}.

Now, we want to present parameters $U=(U_1,U_2,\dots)$  as follows
 \be\label{U-t}
U_n =-\sum_{m=1,3,5,\dots}^\infty n^m t_m
 \ee
 where
$t=(t_1,t_3,t_5,\dots)$ is a new set of parameters (the same
parametrization was used in \cite{nff}).

\br Let us notice that our main example (\ref{r-n-beta}) is not
well fitted into such parametrization. It seems that one needs to
introduce an additional flow parameter
 \[
U_n =-\sum_{m=1,3,5,\dots}^\infty n^m t_m+\beta \ln\; n!
=-\sum_{m=1,3,5,\dots}^\infty n^m t_m+\beta (n\;\ln\; n - n)
 \]
 and consider the discrete dynamics with respect to $\beta$.
 In the large $\time $ limit we may replace $\ln\;n!$ by $n\;\ln\; n -
 n$, then it may be related to the flow introduced in  in \cite{Eguchi}. \er

Then, the right hand side of (\ref{Zz-serie}) may be written as
the following vacuum expectation value (compare with \cite{BHLO})
 \be\label{dualBKPtau}
Z(U;z)=e^{\frac{ z^2}{4}}\langle 0|
e^{H(t)}e^{C(z)}\left(1+\sqrt{2}\sum_{n\ge
0}\frac{z^n}{n!}\phi(n)\phi_0 \right) |0\rangle \ee
 \be\label{C(z)}
 C(z)=\frac 12 \sum_{n,m\ge
0}\frac{z^{n+m}}{n!m!}\phi(n)\phi(m)sign(m-n),\quad
\phi(x):=\sum_{k=-\infty}^{+\infty} x^k\phi_k
 \ee

The proof is similar to the proof in \cite{BHLO} its sketch may be
found in the Appendix \ref{The relation between}.

Now we notice that (\ref{dualBKPtau}) is another example of the
BKP tau function \cite{KvdLbispec} which may be related to the
so-called resonant multi-soliton solution where the number of
solitons is infinite and momentum of solitons are nonnegative
numbers (compare with \cite{LS}). Thus, the tau function of this
(dual) BKP hierarchy is a generating function for normalization
functions. Via (\ref{U-t}), the higher times
$t=(t_1,t_3,t_5,\dots)$ of this dual BKP hierarchy parametrize
hopping rates of the particles of our stochastic model.

Let us we note that time variables $t=(t_1,t_3,t_5,\dots)$ are
integrals of motion for the BKP hierarchy mentioned in (I), in
this sense  the BKP (II) may be referred as a dual to the BKP (I).

At last let us note that the example (\ref{r-n-beta}) where we put
$r=e^{t_1}$ is related to
\[
Z(U;z)=Z(t_1,\beta;z)=e^{\frac{ z^2}{4}}\langle 0|
e^{H_1t_1}e^{C(z)}\left(1+\sqrt{2}\sum_{n\ge
0}\frac{z^n}{(n!)^\beta}\phi(n)\phi_0 \right) |0\rangle
\]
 \[
C(z)=\frac 12 \sum_{n,m\ge
0}\frac{z^{n+m}}{(n!m!)^\beta}\phi(n)\phi(m)sign(m-n),\quad
\phi(x):=\sum_{k=-\infty}^{+\infty} x^k\phi_k
 \]
 One may conjecture that $Z(t_1,\beta;z)$ is a solution to a difference (with
 the respect to $\beta$) -differential (with respect to $t_1$)
 Hirota equation.

\section*{Discussion}

 In the continuation of this paper we
shall consider our model where the parameter $\beta > 2$ describes
a model where a condensate of particles (the region of full
package) fills a region near the origin. We shall explain the
appearance of a phase transition at $\beta=2$ which is rather
similar to the transition studied in \cite{Douglas-Kazakov}. The
other model where the injection rate is a free parameter will be
considered where an analog of the phase transition will be
presented, then it may be interesting to discuss links with
\cite{Schutz}. It is interesting to understand links with results
of \cite{TW} and with the approach of \cite{Forr1}. It may be also
interesting to link our results with the results of a recent paper
\cite{Marshakov} where in the context of the quasiclassical limit
of Toda lattice hierarchy of integrable equations the
Vershik-Kerov limit shape for random partitions was reproduced.

\section*{Acknowledgements}
We are thankful to John Harnad for kind hospitality and numerous
fruitful discussions which allowed to create this paper which may
be viewed as a continuation of \cite{rand} and \cite{BHLO}. We
thank  Marco Bertolla and other participants of the working
seminar on Integrable Systems, Random Matrices and Random
Processes in Concordia university  headed by J. Harnad for
interesting joyful discussions. We thank Anton Zabrodin for a
discussion related to the methods presented in \cite{Gakhov}.

\appendix
\section{Appendices}

\subsection{A remarks on BKP hierarchies \cite{KvdLbispec} and
\cite{DJKM'} and  related vacuum expectation values}\label{A
remarks}

Let us note that different vacuum states were used in the
constructions of BKP hierarchy in versions \cite{KvdLbispec} and
\cite{DJKM'}. If we denote the left and right vacuum states used
in \cite{DJKM'} respectively by $'\langle 0|$ and $|0\rangle'$
then
 \be
\langle 0|=\frac{1}{\sqrt{2}}\; {'\langle 0|}+{'\langle 0|}\phi_0
 ,\qquad
|0\rangle=\frac{1}{\sqrt{2}}|0\rangle' + \phi_0 |0\rangle'
 \ee
Introduce also
 \be '\langle 1|=\sqrt2\; '\langle 0|\phi_0,\qquad
|1\rangle'=\sqrt2 \phi_0|0\rangle', \ee then $'\langle
0||0\rangle'=\, '\langle 1||1\rangle'=1$ and instead of
(\ref{phi-on-vacuum}) we have \be\label{phi-on-vacuum-DJKM}
\phi_n|0\rangle' =\phi_n|1\rangle' =0, \qquad '\langle
0|\phi_{-n}= \, '\langle 1|\phi_{-n}=0,\qquad  n<0
 \ee
 see \cite{DJKM'} for details.

Correspondingly Fock spaces used \cite{KvdLbispec} and
\cite{DJKM'} are different. From the representational point of
view this definition is somewhat more convenient, since each Fock
module remains irreducible for the algebra $B_\infty$ which is the
underlying algebra for KP equations of type B (BKP), see
\cite{KvdLbispec}.

The vacuum states $'\langle 0|$ and $|0\rangle'$ are more familiar
objects in physics. In particular any vacuum expectation value of
an odd number of fermions vanishes, while, for instance,  $\langle
0|\phi_0|0\rangle =\frac{1}{\sqrt{2}}$.

Let $F$ be a product of even number of fermions. Then it is easy
to see that
 \be
\langle 0|F|0\rangle = {'\langle 0|}F{|0\rangle '}
 \ee
Let us note that all vacuum expectation values (v.e.v.)  used in
the present paper (say, tau functions (\ref{BKP-A}) and
(\ref{dualBKPtau})) are sums of v.e.v. of monomials containing
even number of fermions.

 In the context of applications to random process we can construct a
 model of random motion on the semi-infinite lattice based on the
 Fock space used in \cite{DJKM'}, however we find that
 the Fock space used in \cite{KvdLbispec} is much more natural to our point of
view.

\subsection{A remark on formulae containing $Q_\lambda$ functions}\label{A remark on formulae
containing}

Here we shall show that formulae found in \cite{You} are of use in
our case (see also \cite{nff}).

From \cite{You} it is known that
 \be
\begin{split}
{'\langle 0|}e^{H(s)}\phi_{\lambda_1}\phi_{\lambda_2}\cdots\phi_{\lambda_N}|0\rangle'=&
\begin{cases}2^{-\frac{N}2}Q_{(\lambda_1,{\lambda_2},\ldots, {\lambda_N})}(\frac{s}2)
\quad&\mbox{for }N\quad\mbox{even},\\
0&\mbox{for }N\quad\mbox{odd},
\end{cases}\\
\sqrt 2\,  {'\langle 0|}\phi_0e^{H(s)}\phi_{\lambda_1}\phi_{\lambda_2}\cdots
\phi_{\lambda_N}|0\rangle'=&\begin{cases}2^{-\frac{N}2}Q_{(\lambda_1,{\lambda_2},\ldots,
{\lambda_N})}(\frac{s}2)\quad&\mbox{for }N\quad\mbox{odd},\\
0&\mbox{for }N\quad\mbox{even},
\end{cases}\\
\end{split}
 \ee
where $Q_\lambda\left(\frac{s}{2}\right),\;
\lambda=(\lambda_1,\lambda_2,\dots)$ are the projective
Schur functions, see \cite{Mac}.
Thus
\be
\begin{split}
&\langle 0|e^{H(s)}\phi_{\lambda_1}\phi_{\lambda_2}\cdots\phi_{\lambda_N}|0\rangle=\\[3mm]
&\left(\frac{1}{\sqrt{2}} {'\langle 0|}+{'\langle 0|}\phi_0
 \right)e^{H(s)}\phi_{\lambda_1}\phi_{\lambda_2}\cdots\phi_{\lambda_N}
\left(\frac{1}{\sqrt{2}}|0\rangle' + \phi_0 |0\rangle'\right)=\\[3mm]
&\frac12\, {'\langle 0|}e^{H(s)}\phi_{\lambda_1}\phi_{\lambda_2}\cdots\phi_{\lambda_N}|0\rangle' +
{'\langle 0|}\phi_0e^{H(s)}\phi_{\lambda_1}\phi_{\lambda_2}\cdots\phi_{\lambda_N}\phi_0|0\rangle' +\\
&+\frac{1}{\sqrt 2}\, {'\langle 0|}\phi_0 e^{H(s)}
\phi_{\lambda_1}\phi_{\lambda_2}\cdots\phi_{\lambda_N}|0\rangle'+
\frac{1}{\sqrt 2}\, {'\langle 0|}e^{H(s)}\phi_{\lambda_1}\phi_{\lambda_2}\cdots\phi_{\lambda_N}\phi_0|0\rangle'\,
=\\[3mm]
&\frac12\, {'\langle 0|}e^{H(s)}\phi_{\lambda_1}\phi_{\lambda_2}\cdots\phi_{\lambda_N}|0\rangle' +\frac12\,
{'\langle 1|}e^{H(s)}\phi_{\lambda_1}\phi_{\lambda_2}\cdots\phi_{\lambda_N}|1\rangle' +\\
&+\frac{1}{\sqrt 2}\, {'\langle 0|}\phi_0 e^{H(s)}
\phi_{\lambda_1}\phi_{\lambda_2}\cdots\phi_{\lambda_N}|0\rangle'+
\frac{1}{\sqrt 2}\, {'\langle 1|}\phi_0 e^{H(s)}\phi_{\lambda_1}\phi_{\lambda_2}\cdots\phi_{\lambda_N}|1\rangle'\, =\\[3mm]
&2^{-\frac{N}2}Q_{(\lambda_1,{\lambda_2},\ldots, {\lambda_N})}\left(\frac{s}2\right)\, ,
\end{split}
\ee
since the role of ${'\langle 0|}$ and ${'\langle 1|}$ (resp.
$|0\rangle'$ and $|1\rangle'$) is interchangeable.

\subsection{Case $0<\beta<2 $: Vershik-Kerov type
asymptotic}\label{Case}

We want to solve the following singular integral equation
 where $r$ is a given constant:
\be\label{int-sing-eq-A4} \ln \frac{r}{\lambda^{\beta}} +P
\int_0^R \frac{\sigma(xR^{-1})dx}{\lambda-x}- P \int_0^R
\frac{\sigma(xR^{-1})dx}{\lambda+x}+\frac12 \ln 2\left(
\time-\int_{0}^{R}x\sigma(x R^{-1})dx \right)=0
 \ee
where $\lambda\in[0,R]$ and where $\sigma$ is defined on the
interval $x\in [0,R]$. Here and below  $P$ serves to denote the
principal value of integrals.

\br Let us notice that the second term in (\ref{int-sing-eq-A4})
describes the repulsion of charges distributed with density
$\sigma$ along $(0,R)$, while the third term may be interpreted as
the attraction of these charges to their image in the mirror.
Then, it is natural to continues $\sigma$ to the interval $(-R,0)$
such that $\sigma(-x):=-\sigma(x)$ (this describes the replacing
of particles (with coordinate, say, $x$) by holes (situated in
$-x$).
 \[
P \int_0^R \frac{\sigma(xR^{-1})dx}{\lambda-x}- P \int_0^R
\frac{\sigma(xR^{-1})dx}{\lambda+x}=P \int_{-R}^R
\frac{\sigma(xR^{-1})dx}{\lambda-x}
 \]

Notice that in this case in general we get a jump of $\sigma$ in
$x=0$. For our future purpose (of inverting Hilbert type
transforms) we prefer to deal with continues functions. For this
purpose we shall modify $\sigma$ in the region $(-R,0)$ by adding
a constant equal to this jump, $\Delta$, getting continues
modified $\sigma$ (compensating this change of $\sigma$ by adding
logarithmic term to the right hand side), see below \er

 Now, $\sigma(x)$ is a function defined on the whole
interval $[-R,R]$ via relation
 \be\label{symmetry-condition}
\sigma(-x)=\Delta - \sigma(x),\quad x\in [-R,R]
 \ee
(As we shall see later $\Delta=\beta $ if $0 < \beta < 2$, and
$\Delta=2$ if $\beta > 2$). Then , we re-write
(\ref{int-sing-eq-A4}) in a more convenient form as
 \be\label{Hilbert-on-interval}  P
\int_{-R}^R \frac{\sigma(xR^{-1})dx}{\lambda-x}=\ln
(R+\lambda)^{\Delta} +\ln\lambda^{\beta-\Delta} -\ln (Cr),\quad
\lambda\in[-R,R]
 \ee
 where we denoted
 \be\label{C}
C^2:= 2\left(
\time-\int_{0}^{R}x\sigma(x R^{-1})dx \right)
 \ee
 $C$  is an independent of $\lambda$  constant to be defined later.

 Singular integral equation (\ref{Hilbert-on-interval}) may be solved by standard
 method, see \cite{Gakhov}.
 In case $\sigma(xR^{-1})$ is continues and bounded on
 $[-R,R]$ the solution is given by formulae
 \be\label{general-solution}
\sigma(xR^{-1})=-\pi^{-2} P\int_{-R}^{R}
\frac{\sqrt{x^2-R^2}}{\sqrt{\lambda^2-R^2}}
\ln\frac{(\lambda+R)^{\Delta}\lambda^{\delta}}{C r} \frac{d
\lambda}{x-\lambda},\quad x \in [-R,R],
 \ee
($\delta:=\beta -\Delta$) see formula (42.26) in \cite{Gakhov}.
Let us evaluate the integral in an explicit way. We consider
$\sqrt{\lambda^2-R^2}$ in the integrand of
(\ref{general-solution}) as single-valued function with the cut
$[-R,R]$ whose upper limit on $[-R,R]$ is positive and lower limit
on $[-R,R]$ is negative. Also we shall consider $\ln
(\lambda+R)^{\Delta}$ in the integrand as the single valued
function defined on the whole complex plane with the cut
$[-\infty,-R]$. The cut of $\ln \lambda^\delta$ will be viewed on
the ray $(-\infty,0)$,  a little bit above the real axe. Then we
have
 \be\label{complex-plane-integral}
 P\int_{-R}^{R}
\frac{\sqrt{x^2-R^2}}{\sqrt{\lambda^2-R^2}}
\ln\frac{(\lambda+R)^{\Delta}\lambda^\delta}{C r} \frac{d
\lambda}{x-\lambda}= -\frac 12 \oint_{C}
\frac{\sqrt{x^2-R^2}}{\sqrt{\lambda^2-R^2}}
\ln\frac{(\lambda+R)^{\Delta}\lambda^\delta}{C r} \frac{d
\lambda}{x-\lambda}
 \ee
where the contour $C$ is going crossing the points $0$ and $2R$:
$C=-C_++C_-$. One can inflate the contour through the point $2R$
as there are no 'bad' singularities there. We have a cut
$(-\infty,0)$ caused by the logarithm. Inflating the contour to
the infinity we see that the only contribution will be caused by
the cut of the logarithm (we come to the contour which starts on
minus infinity going a little bit upper the real line, than
turning at the origin and going back to minus infinity a little
bit under the real line. (The contribution of the circle embracing
infinity vanishes as the integrand's asymptotic is ${d\lambda
\over \lambda^2}$.) This yields the integral along the cut
 \be
 P\int_{-R}^{R}
\frac{\sqrt{x^2-R^2}}{\sqrt{\lambda^2-R^2}}\ln\frac
{(\lambda+R)^{\Delta}\lambda^\delta}{C r} \frac{d
\lambda}{x-\lambda} =
 \ee
 \be\label{along-cut}
-\frac 12\int_{-R}^0
\frac{\sqrt{x^2-R^2}}{\sqrt{\lambda^2-R^2}}(2\pi \delta\sqrt{-1})
\frac{d \lambda}{x-\lambda}
 -\frac 12\int_{-\infty}^0
\frac{\sqrt{y^2-2Ry}}{\sqrt{h^2-2Rh}}(2\pi \Delta\sqrt{-1})
\frac{d h}{y-h}
 \ee
 where $h=\lambda +R$, $y=x+R $ and $2\pi \Delta\sqrt{-1}$ and $2\pi \Delta\sqrt{-1}$ are
  the jumps of the logarithm.

The first integral is
\cite{Brychkov-Prudnikov}
 \[
\int_{-R}^0 \frac{\sqrt{x^2-R^2}}{\sqrt{\lambda^2-R^2}}(2\pi
\delta\sqrt{-1}) \frac{d \lambda}{x-\lambda}=\arcsin\frac{-R}{2x}
 \]
As we see taking small $x$ this integral is not a real number; this results in
the condition $\delta=0$, or, the same
 \[
\Delta=\beta
 \]
  For the second integral of (\ref{along-cut}) we have (see \cite{Brychkov-Prudnikov}):
  \be\label{Brychkov-Prudnikov}
\int_0^{+\infty}\frac{1}{\sqrt{h^2+2Rh}} \frac{d
h}{y+h}=\frac{1}{\sqrt{-y^2+2Ry}}\arccos \left( \frac yR -1
\right)
  \ee
 Inserting the last formula into (\ref{along-cut})
  and then into (\ref{general-solution}) we finally obtain
  \be\label{alpha-sigma-A4}
\sigma(xR^{-1})=\frac {\beta}\pi \arccos \frac xR,\quad x\in
[0,R],\quad 0\le \beta \le 2
  \ee
We add the last un-equality to provide (\ref{+2}), $0\le\sigma\le 1$.

At last let us mark that one may show that if we relate the
asymptotic configuration (\ref{alpha-sigma-A4}) to a strict
partition, then, the shape of the Young diagram of the double of
this partition will coincide with the so-called Vershik-Kerov
asymptotic shape \cite{Kerov-Vershik},\cite{Kerov-Vershik-2}.

\subsection{Two useful integrals}\label{Two useful integrals}

One can show that
 \be
 \frac {1}{\pi} \int_0^1 \arccos u \; \left(u\ln
u-u \right)du =-\frac 18 -\frac{\ln\; 2}{8}
 \ee
 \be
  \frac {1}{\pi^2}
  \int_0^1 du \int_0^u \arccos u \; \arccos u'\;
\ln\;\frac{u-u'}{u+u'}\; du'=-\frac{1}{16}
 \ee
which yields $A = \frac{1}{16}+\frac{\ln\; 2}{8} $.

\subsection{The relation between formula (\ref{dualBKPtau}) and (\ref{Zz-serie})}\label{The relation between}

Recal from (\ref{C(z)}) that $\phi(z)=\sum_{i\in\mathbb{Z}}z^i\phi_i$. Using (\ref{canonical})
and (\ref{phi-on-vacuum}), we can also calculate the vacuum expectation value for two of these fields:
\begin{equation}
\begin{split}
\label{wickphi}
  \langle 0|\phi(z_1)\phi(z_2) |0\rangle=&
\sum_{m>0}(-z_2/z_1)^m+\tfrac12
 \\[3mm]
  =&\frac{1}2\frac{z_1-z_2}{z_1+z_2}\, ,
 \end{split}
\end{equation}
(where we assume that $|z_1|>|z_2|$). Using Wick's Theorem we obtain that
\begin{equation}
\label{Wickfields}
\begin{split}
\langle 0|\phi(z_1)\phi(z_2)\cdots\phi(z_{2m})|0 \rangle=&
\left(\frac12\right)^m
\prod_{1\le i< j\le 2m} \frac{
  z_i-z_j}{
  z_i+z_j}
\, ,
\\[3mm]
\langle 0| \phi(z_1)\phi(z_2)\cdots\phi(z_{2m+1})\phi_0|0 \rangle=&
\left(\frac12\right)^{m+1}\prod_{1\le i< j\le 2m+1} \frac{
  z_i-z_j}{
  z_i+z_j}
\, .
\end{split}
\end{equation}
For the exponentials of the Hamiltonians $H(t)$ and $\bar H(t)$ introduced in
(\ref{Hamiltonians}) one has
\be
\label{Ham-on-vac}
e^{H(t)}| 0\rangle=|0\rangle\qquad \langle 0|e^{\bar H(t)}=\langle 0|
\ee
and
\be
\label{Ham-on-phi}
e^{H(t)}\phi(z)e^{-H(t)}=b(t,z)\phi(z),\qquad
e^{\bar H( t)}\phi(z)e^{-\bar H( t)}=b( t,z^{-1})\phi(z),
\ee
where
\be
b(t,z)=\exp\left( \sum_{n\ge 1,\;\text{odd}}t_{ n}z^n\right)
\ee
Using (\ref{Wickfields}), (\ref{Ham-on-vac}) and (\ref{Ham-on-phi}) we calculate
\begin{equation}
\label{dualBKPtau1}
\begin{split}
& 2^m\sum_{{\rm all}\ x_i>0}\langle
0|e^{H(t)}\phi(x_1)\phi(x_2)\cdots\phi(x_{2m})|0\rangle
\prod_{i=1}^m sign(x_{2i-1}-x_{2i})
\\
=& \sum_{{\rm all}\ x_i>0} \prod_{k=1}^{2m}b(t,x_k) \prod_{1\le
i<j\le 2m}\frac{x_i-x_j}{x_i+x_j} \prod_{i=1}^m
sign(x_{2i-1}-x_{2i})
\\
=& \sum_{\pi\in S_{2m}}\sum_{ x_{\pi(2m)}< x_{\pi(2m-1)}<\cdots<
x_{\pi(1)}} \prod_{k=1}^{2m}b(t,x_k) \prod_{1\le i<j\le
2m}\frac{x_i-x_j}{x_i+x_j} \prod_{i=1}^m sign(x_{2i-1}-x_{2i})
\\
=& \sum_{\pi\in S_{2m}}\sum_{ x_{2m}< x_{2m-1}<\cdots < x_{1}}
sign(\pi^{-1}) \prod_{k=1}^{2m}b(t,x_k) \prod_{1\le i<j\le
2m}\frac{x_i-x_j}{x_i+x_j} \prod_{i=1}^m
sign(\pi^{-1}(2i)-\pi^{-1}(2i-1))
\\
=& 2^m m!\sum_{x_{2m}< x_{2m-1}<\cdots
<x_{1}}\prod_{k=1}^{2m}b(t,x_k) \prod_{1\le i<j\le
2m}\frac{x_i-x_j}{x_i+x_j}
\\
=& \frac{2^m m!}{(2m)!}\sum_{{\rm all}\
x_i>0}\prod_{k=1}^{2m}b(t,x_k) \prod_{1\le i<j\le 2m}\left|
\frac{x_i-x_j}{x_i+x_j}\right|
\end{split}
\end{equation}
We have used that
\begin{equation}
\label{pfaff1} \sum_{\pi\in S_{2m}}sign(\pi^{-1}) \prod_{i=1}^m
sign(\pi^{-1}(2i)-\pi^{-1}(2i-1))=2^mm!
\end{equation}
which is obvious if one realizes that the right-hand side of
(\ref{pfaff1}) is equal to $2^m m!$ times the Pfaffian of the
$2m\times 2m$ matrix $A=\left(A_{ij}\right)_{1\le i,j\le 2m}$
where $A_{ij}=sign(j-i)$.

Next, we calculate
\begin{equation}
\label{dualBKPtau2}
\begin{split}
& 2^{m+1}\sum_{{\rm all}\ x_i>0}\langle
0|e^{H(t)}\phi(x_1)\phi(x_2)\cdots\phi(x_{2m})\phi(x_{2m+1})\phi_0|0\rangle
\prod_{i=1}^m sign(x_{2i-1}-x_{2i})
\\
=& \sum_{{\rm all}\ x_i>0} \prod_{k=1}^{2m+1}b(t,x_k) \prod_{1\le
i<j\le 2m+1}\frac{x_i-x_j}{x_i+x_j} \prod_{i=1}^m
sign(x_{2i-1}-x_{2i})
\\
=&
\sum_{\pi\in S_{2m+1}}\sum_{ x_{\pi(2m+1)}< x_{\pi(2m)}<\cdots<
x_{\pi(1)}} \prod_{k=1}^{2m+1}b(t,x_k) \prod_{1\le i<j\le
2m+1}\frac{x_i-x_j}{x_i+x_j} \prod_{i=1}^m sign(x_{2i-1}-x_{2i})
\\
=& \sum_{\pi\in S_{2m+1}}\sum_{ x_{2m+1}< x_{2m}<\cdots < x_{1}}
sign(\pi^{-1}) \prod_{k=1}^{2m+1}b(t,x_k) \prod_{1\le i<j\le
2m+1}\frac{x_i-x_j}{x_i+x_j} \prod_{i=1}^m
sign(\pi^{-1}(2i)-\pi^{-1}(2i-1))
\\
=&\sum_{\ell=1}^{2m+1} \sum_{\pi\in S_{2m+1},\atop
\pi(\ell)=2m+1}\sum_{ x_{2m+1}< x_{2m}<\cdots < x_{1}}
sign(\pi^{-1}) \prod_{k=1}^{2m+1}b(t,x_k) \prod_{1\le i<j\le
2m+1}\frac{x_i-x_j}{x_i+x_j}
\prod_{i=1}^m sign(\pi^{-1}(2i)-\pi^{-1}(2i-1))\\
=& \sum_{\rho\in S_{2m}}\sum_{ x_{2m+1}< x_{2m}<\cdots < x_{1}}
sign(\rho) \prod_{k=1}^{2m+1}b(t,x_k) \prod_{1\le i<j\le
2m}\frac{x_i-x_j}{x_i+x_j}
\prod_{i=1}^m sign(\rho(2i)-\rho(2i-1))\\
=& 2^m m!\sum_{x_{2m+1}< x_{2m}<\cdots
<x_{1}}\prod_{k=1}^{2m+1}b(t,x_k) \prod_{1\le i<j\le
2m+1}\frac{x_i-x_j}{x_i+x_j}
\\
=& \frac{2^m m!}{(2m+1)!}\sum_{{\rm all}\
x_i>0}\prod_{k=1}^{2m+1}b(t,x_k) \prod_{1\le i<j\le 2m+1}\left|
\frac{x_i-x_j}{x_i+x_j}\right|
\end{split}
\end{equation}
Here we have used that
\[
\begin{split}
\sum_{\pi\in S_{2m+1},\atop \pi(\ell)=2m+1} sign(\pi^{-1})
\prod_{i=1}^m sign(\pi^{-1}(2i)-\pi^{-1}(2i-1))&=\\
-(-)^\ell \sum_{\nu\in S_{2m+1},\atop \nu(\ell)=\ell}sign(\nu)
\prod_{i=1}^m  sign(\nu(\epsilon_\ell(2i))-\nu(\epsilon_\ell(2i-1)))&=\\
-(-)^\ell \sum_{\rho\in S_{2m}}sign(\rho) \prod_{i=1}^m
sign(\rho(2i)-\rho(2i-1))\, ,
\end{split}
\]
where $\epsilon_\ell(i)=i$ if $i<\ell$ and $\epsilon_\ell(i)=i+1$
if $i>\ell$.
So if
\[
\pi^{-1}=\begin{pmatrix}
1&2&\cdots&\ell-1&\ell&\ell+1&\cdots&2m&2m+1\\j_1&j_2&
\cdots&j_{\ell-1}&j_\ell&j_{\ell+1}&\cdots&j_{2m}&\ell\end{pmatrix}\, ,
\]
then
\[
\nu=\begin{pmatrix}
1&2&\cdots&\ell-1&\ell&\ell+1&\ell+2&\cdots&2m&2m+1\\j_1&j_2&
\cdots&j_{\ell-1}&\ell&j_\ell&j_{\ell+1}&\cdots&j_{2m-1}&j_{2m} \end{pmatrix}
\]
and
\[
\rho=\begin{pmatrix}
1&2&\cdots&\ell-1&\ell&\ell+1&\cdots&2m-1&2m\\i_1&i_2&
\cdots&i_{\ell-1}&i_\ell&i_{\ell+1}&\cdots&i_{2m-1}&i_{2m} \end{pmatrix}\, ,\quad i_k=
\begin{cases}j_k&{\rm if}\ j_k<\ell,\\
j_k-1&{\rm if}\ j_k>\ell,
\end{cases}
\]
For instance for $\ell=3$ and $m=2$ one has
\[
\pi^{-1}=
\begin{pmatrix}
1&2&3&4&5\\ 5&1&4&2&3
\end{pmatrix}
\to \nu=\begin{pmatrix} 1&2&3&4&5\\ 5&1&3&4&2
\end{pmatrix}\to \rho=\begin{pmatrix}
1&2&3&4\\ 4&1&3&2\end{pmatrix}
\]
Hence, if we develop (\ref{dualBKPtau}) as a power series in $z$
and keep the term $e^{\frac{z^2}4}$. Then substitute
(\ref{dualBKPtau1}) and (\ref{dualBKPtau2}) and use (\ref{U-t}),
we obtain (\ref{Zz-serie}).

\end{document}